\documentstyle[aps,twocolumn,psfig,prl]{revtex}

\newcommand{\tilV}{{\tilde{V}}}
\newcommand{\tilt}{{\tilde{t}}}

\newcommand{\tiltau}{{\tilde{\tau}}}

\newcommand{\tilL}{{\tilde{L}}}
\newcommand{\tell}{{\tilde{\ell}}}
\newcommand{\terr}{{\tilde{r}}}
\newcommand{\avell}{\overline{\ell}}
\newcommand{\averr}{\overline{r}}
\newcommand{\Orderpar}{{\mathcal{L}}}

\def \order#1 {${\mathcal{O}}(#1)$\ }

\begin{document}

\def \recent {watts,herzel,pandit,bart,barrat,monasson,newman,weigt,firstorder,kasturi,newman2,kulkarni,corrlen}

\draft

\title{Spreading and shortest paths in systems with sparse long-range
  connections.}

\author{Cristian F.~Moukarzel}

\address{ Instituto de F\'{\i}sica, Universidade ~Federal Fluminense, CEP
  24210-340, Niter\'oi, RJ, Brazil}

\maketitle

\begin{abstract}
  Spreading according to simple rules (e.g. of fire or diseases) and
  shortest-path distances are studied on $d$-dimensional systems with a small
  density $p$ per site of \emph{long-range connections} (``Small-World''
  lattices). The volume $V(t)$ covered by the spreading quantity on an
  infinite system is exactly calculated in all dimensions. We find that $V(t)$
  grows initially as $\Gamma_d t^d/d$ for $t<< t^* = (2p \Gamma_d
  (d-1)!)^{-1/d}$ and later exponentially for $t>>t^*$, generalizing a
  previous result in one dimension.  Using the properties of $V(t)$, the
  average shortest-path distance $\ell(r)$ can be calculated as a function of
  Euclidean distance $r$. It is found that $\ell(r) \sim r$ for $r < r_c = (2
  p \Gamma_d (d-1)!)^{-1/d} \log(2 p \Gamma_d L^d)$ and $\ell(r) \sim r_c$ for
  $r > r_c$.  The characteristic length $r_c$, which governs the behavior of
  shortest-path lengths, \emph{diverges} with system size for all $p>0$.
  Therefore the mean separation $s \sim p^{-1/d}$ between shortcut-ends is not
  a relevant internal length-scale for shortest-path lengths. We notice
  however that the globally averaged shortest-path length $\avell$ divided by
  $L$ is a function of $L/s$ only.
\end{abstract}


Regular $d$-dimensional lattices with a small density $p$ per site of
long-ranged bonds (or ``small-world'' networks)~\cite{watts} model the effect
of weak unstructured (mean-field) interactions in a system where the dominant
interactions have a regular $d$-dimensional structure, and thus may have many
applications in physics as well as in other sciences~\cite{\recent}.

In this work we study the spreading (See e.g.~\cite{watts,newman2} and
references therein) of some influence (e.g. a forest fire, or an infectious
disease) according to the following simple law: we assume that, at each
time-step, the fire or disease propagates from an burnt (or infected) site to
all unburnt (uninfected) sites connected to it by a link.  Long-range
connections, or \emph{shortcuts} represent sparks that start new fires far
away from the original front, or, in the disease-spreading case, people who
when first infected move to a random location amongst the non-infected
population.  For the dynamics of this simple problem, an important network
property is the set of shortest-path distances $\{\ell_{ij}\}$, where
$\ell_{ij}$ is defined as the minimum number of links one has to traverse
between $i$ and $j$.  On isotropic $d$-dimensional lattices $\ell_{ij}$ is
proportional to $d^E_{ij}$, the Euclidean distance between $i$ and $j$. On
regular lattices, both the number of sites within an Euclidean distance $r$
from $i$, and the number of sites within $r$ nearest-neighbor steps from $i$
behave as $r^d$.

Disorder can modify this in several ways. On a random fractal, the number of
sites contained in a volume of Euclidean radius $r$ is
$r^{d_f}$~\cite{Stauffer,Fractals}, where $d_f < d$ is the fractal dimension.
On the other hand, the number of sites visited in at most $\ell$ steps is
$\ell^{d_f / d_{min}}$~\cite{Stauffer,Fractals}, where $d_{min} \geq 1$ is the
\emph{shortest-path dimension}, defined such that $\ell \sim r^{d_{min}}$.  On
fractals, shortest-path lengths $\ell(r)$ are thus much \emph{larger} than
Euclidean distances $r$.

Consider now a randomly connected network. If we have $L^d$ sites sitting on a
regular $d$-dimensional lattice, but connect them at random with an average
coordination number $C$ (i.e. a total of $L^d C/2$ bonds), the number of sites
in a volume of radius $r$ is still $r^d$, but we can visit $\sim C^k$ sites in
$k$ steps. Thus all $L^d$ sites can be visited in \order{{\log L^d}} steps,
and therefore the typical shortest-path distance $\avell$ is of order $\log
L$, much \emph{shorter} than the typical Euclidean distance $\averr \sim L$.

``Small-World'' networks~\cite{watts} are intermediate between the regular
lattice, where $\avell \sim L$, and the random graph, where $\avell \sim \log
L$. They consist of a regular $d$-dimensional lattice with $N=L^d$ sites, on
which $p L^d$ \emph{additional} long-range bonds have been connected between
randomly chosen pairs of sites.  The key finding of Watts and
Strogatz~\cite{watts} is that a vanishingly small density $p$ of long-range
bonds is enough to make shortest-path distances proportional to $\log L$
instead of $L$. If $L^d p << 1$, the system typically contains no shortcuts,
and the average shortest-path distance $\avell=1/N^2\sum_{<ij>} \left
  [\ell_{ij}\right ]_p$ scales as $L$. If on the other hand $L^d p >>1$, one
finds $\avell \sim \log{L}$~\cite{watts,bart,newman}. For any fixed density
$p$ of long-ranged bonds, a \emph{crossover-size} $L^*(p)$
exists~\cite{bart,barrat}, above which shortest-path distances are only
logarithmically increasing with $L$. This crossover size diverges as
$p^{-1/d}$~\cite{bart,newman,firstorder} when $p \to 0$. The nature of the
small-world transition at $p=0$ was recently discussed
controversially\cite{newman,firstorder,newman2,corrlen}. It is still a matter
of debate~\cite{newman2,corrlen} whether $p=0$ can be regarded as being
equivalent to a critical point.

In this work we calculate the volume $V(t)$ that is covered, on a small-world
network, by a spreading quantity as a function of time~\cite{newman2} when the
spreading law is the simple rule above, and derive an exact expression for the
average shortest-path $\ell(r)$ on these systems.

\begin{figure}[thpb]
\centerline{\psfig{figure=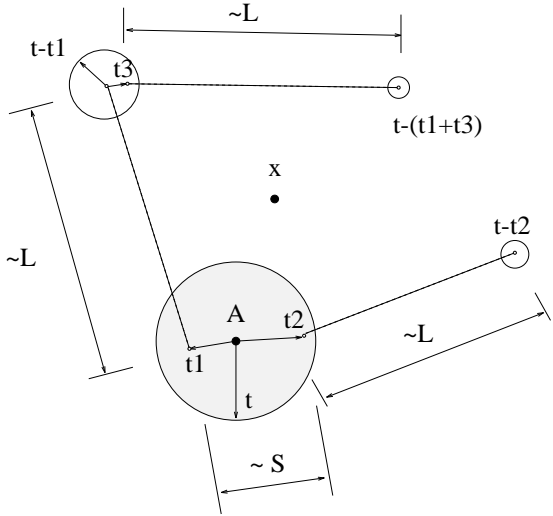,width=8cm,angle=270}}
\caption{ {} Consider the spreading of fire or diseases on small-world
  systems. Assume that $\rho L^d$ points are drawn at random in
  $d$-dimensional space (open dots). The mean distance $S$ between neighboring
  points is proportional to $\rho^{-1/d}$. Now connect pairs of points at
  random.  The mean separation between ``mates'' will be of the order of $L$,
  the system size. Paired points represent long-range bonds (shaded lines),
  across which fire or diseases travel instantaneously. Now a disease starts
  to spread from $A$.  Each time that the resulting sphere hits a
  shortcut-end, a secondary sphere will be born somewhere else. These in turn
  later give rise to other secondary spheres. The proliferation of secondary
  spheres produces an exponentially fast growth of the infected volume for
  times $t>S$.  }
\label{fig:1}
\end{figure}

Assume a disease spreads with constant radial velocity $v=1$ from an original
infection site $A$, as shown in Fig.~\ref{fig:1}. Let $\rho = 2p$ be the
density of \emph{shortcut-ends} on the system. We work on the continuum for
simplicity, so that the infected volume will initially grow as a sphere of
radius $t$ and surface $\Gamma_d t^{d-1}$. We call the sphere stemming from
$A$ ``primary sphere''.

Each time the primary sphere hits a shortcut end, which happens with
probability $\rho \Gamma_d t^{d-1}$ per unit time, a new sphere
(``secondary'') starts to grow from a random point in non-infected space (the
other end of the shortcut). These in turn later give rise to further secondary
spheres in the same fashion.

Following Newman and Watts~\cite{newman2}, we notice that the total infected
volume is the sum of the primary volume $\Gamma_d \int_0^t \tau^{d-1} d\tau$
plus a contribution $V(t-\tau)$ for each new sphere born at time $\tau$. Thus
in the continuum the average total infected volume satisfies
\begin{equation}
V(t)= \Gamma_d \int_0^t \tau^{d-1} \left\{ 1 + \rho V(t-\tau) \right\} d\tau,
\label{eq:original}
\end{equation}
which can be rewritten in terms of rescaled variables $\tilV=\rho V$ and
$\tilt=(  \rho \Gamma_d (d-1)!)^{1/d} t$ as
\begin{equation}
\tilV(\tilt)= \frac{1}{(d-1)!} \int_0^{\tilt}
( \tilt- \tiltau )^{d-1} 
\left\{ 1 + \tilV(\tiltau) \right\} d\tiltau
\label{eq:V}
\end{equation}

It is interesting to notice that $\tilV$ is the total number of infected
shortcut-ends, while $\tilt^d/d!$ is the total number of shortcut-ends infected
by the primary sphere. On an infinite system, the functional relation
(\ref{eq:V}) that links these two variables has no parameters except for the
space dimensionality $d$. On a system of finite volume $\Gamma_d L^d/d$, an
important parameter is the rescaled linear size $\tilL = (  \rho
\Gamma_d (d-1)!)^{1/d}L$, whose $d$-th power gives the total number $N_s$ of
shortcut-ends as $N_s = \tilL^d/d!$.

Deriving (\ref{eq:V}) $d$ times with respect to $\tilt$ we obtain 
\begin{equation}
\frac{\partial^d }{\partial \tilt^d}V(\tilt)= 1 + V(\tilt),
\end{equation}
whose solution is 
\begin{equation}
\tilV(\tilt)= \sum_{k=1}^{\infty} 
\frac{ \tilt^{dk} }{(dk)!} 
\label{eq:final}
\end{equation}

Notice that (\ref{eq:final}) is a series expansion of $(e^{\tilt}-1)$ with all
powers not multiples of $d$ removed. Thus (\ref{eq:final}) can be written as a
sum of $d$ exponentials, each with a different $d$-root of $1$ in its
argument. In this way, powers which are not multiples of $d$ cancel out.
\begin{equation}
\tilV(\tilt)= \frac{1}{d} \sum_{n=0}^{d-1} \exp\{\mu_d^n
\ \tilt \} -1
\end{equation}
where $\mu_d=e^{i 2\pi/d}$. Some specific examples are
\[\begin{array}{lcll}
\tilV(\tilt)&=& e^{\tilt}-1  &\hbox{in 1d,} \\ \\
\tilV(\tilt)&=& \cosh{\tilt}-1 &\hbox{in 2d,} \\ \\
\tilV(\tilt)&=& 
\frac{ e^{\tilt}+e^{{\mu_3}\tilt}+ e^{{\mu_3}^2\tilt}}{3}-1
&\hbox{in 3d,} 
\end{array}\]
where the one-dimensional solution coincides with that previously derived by
other methods~\cite{newman2}.

A general property of (\ref{eq:final}) is that $\tilV$ grows as $\tilt^d/d!$
for $\tilt < 1$, and later exponentially as $e^{\tilt}/d$. Thus the
characteristic timescale~\cite{newman2} for the spreading process is $t^* =
(\rho \Gamma_d (d-1)!)^{-1/d}$.

Notice that (\ref{eq:original}), and thus also (\ref{eq:final}), only hold on
an infinite system.  On a finite system with $\tilL^d >>1$, $\tilV$ will
saturate after a time $\tilt_{sat}$ that can be estimated by equating $\tilV
\sim e^{\tilt_{sat}}/d \sim \tilL^d/d!$ and therefore
\begin{equation}
\tilt_{sat} \sim \log(\tilL^d/(d-1)!),
\end{equation}
which can be rewritten as
\begin{equation}
t_{sat} \sim (\rho \Gamma_d (d-1)!)^{-1/d}
\log(\rho \Gamma_d  L )
\label{eq:tsat}
\end{equation}
If on the other hand $\tilL^d <<1$, the spreading stops at $\tilt_{sat} =
\tilL$, before reaching the exponential growth regime.

Thus for a finite system with $\tilL^d >>1$ one has
\begin{equation}
\tilV(\tilt) d \sim \left\{
\begin{array}{lcl}
  \tilt^d/d! &\hbox{ for }& \tilt << 1 \\
  e^{\tilt}/d &\hbox{ for }& 1<< \tilt < \tilt_{sat} \sim
  \log(\frac{\tilL^d}{(d-1)!}) 
  \\ 
  \tilL^d/d! &\hbox{ for }& \tilt > \tilt_{sat}
\end{array}
\right.
\end{equation}

Assume now that $\tilL^d >>1$.  Because of the exponentially fast spreading
process, the fraction of the total volume covered by the disease is negligible
for $\tilt < \tilt_{sat}$ and saturates to one abruptly at $t=t_{sat}$.
Therefore on a large system most of the points become infected essentially at
the same time $\tilt_{sat}$.

Now let us see how to calculate the average shortest-path distance $\ell(r)$
as a function of the Euclidean separation $r$ between two points. Since we
assumed that the disease spreads with unit velocity, it is clear that the time
$t$ at which a point $x$ becomes first infected is exactly the shortest-path
distance $\ell(A,x)$ from $A$ to $x$. By definition, no part of the finite
system remains uninfected after $t=t_{sat}$, so we conclude that no
shortest-path distance can be larger than $t_{sat}$ on a finite system.
Combining this with the fact that $\ell(r)$ cannot decrease with increasing
$r$, we conclude that
\begin{equation}
\ell(r) = t_{sat}\qquad \qquad \hbox{for} \ r \geq t_{sat}
\label{eq:saturates}
\end{equation}

In order to calculate $\ell(r)$ for $r < t_{sat}$, let us write $V(t) = V_1(t)
+ V_2(t)$, where $V_1$ is the primary volume and $V_2$ the volume infected by
secondary spheres.

Let $p_2(t)$ be the probability to become infected by the secondary infection
exactly at time $t \leq t_{sat}$. Consequently $I_2(t)=\int_0^t p_2(\tau)
d\tau$ is the probability for a point to become infected at time $t$ or
earlier.  Assuming that $p_2(t)$ is known, it is easy to calculate the average
shortest-path distance $\ell(r)$ as a function of Euclidean distance $r$,
according to the following. If an individual at $x$ becomes infected by a
secondary sphere at time $\tau < d^E(A,x)$, its shortest-path distance
$\ell(A,x)$ to $A$ is $\tau$. Otherwise if $x$ is still uninfected at time
$t=d^E(A,x)$ (which happens with probability $1 - I_2(d^E(A,x))$), then
$\ell(A,x)=d^E(A,x)$, since at that time the primary sphere hits $x$ with
probability one. Therefore the average shortest-path satisfies
\begin{eqnarray}
  \ell(r) &=& \int_0^{r} t p_2(t) dt + r  \left\{1 - I_2(r)\right\} \\ 
  &=& r -  \int_0^{r} I_2(t) dt 
  \label{eq:ell1}
\end{eqnarray}

The fact that the secondary volume $V_2$ is randomly distributed in space
makes this problem relatively simple. The probability $I_2(t)$ for a point to
be infected by the secondary version of the disease at time $t$ or earlier is
simply $I_2 = V_2(t)/(1/d \Gamma_d L^d)$, i.e. the fraction of the total
volume which is covered by the secondary infection. Thus
\begin{equation}
  \ell(r) = r -  \frac{d}{ \Gamma_d L^d} \int_0^{r} V_2(t)dt
  \label{eq:ell2}
\end{equation}

If there are no shortcuts on the system, $V_2$ is zero at all times and thus
$\ell(r)=r$ as expected. But it is also clear from this expression that
$\ell(r) = r$ when $L \to \infty$, for all \emph{finite} $r$, i.e. in the
thermodynamic limit the shortest-pat distance $\ell(r)$ coincides with the
Euclidean distance $r$ for all \emph{finite} $r$, no matter what $\rho$ is.

On a finite system with $\tilL^d>>1$, $V_2(t)/L^d$ is negligible for all $t <
t_{sat}$ as we have already noticed. Therefore $\ell(r)=r$ if $r < t_{sat}$.
Combining this with (\ref{eq:saturates}) we have
\begin{equation}
  \ell(r) \approx \left\{
    \begin{array}{lll}
      r &\hbox{for}  &r < r_c = (\rho \Gamma_d (d-1)!)^{-1/d} \log(\rho
      \Gamma_d L^d) \\ 
      r_c &\hbox{for}  &r \geq r_c 
    \end{array} \right.
\label{eq:ellfinal}
\end{equation}
Detailed knowledge of $\ell(r)$ for $r = r_c$ would only be possible if
the finite-size effects that we ignored in (\ref{eq:V}) were exactly known,
but the interesting remark is that the lack of this knowledge has little or no
importance for $r \neq r_c$.

We thus see that on a \emph{finite} system, a characteristic length $r_c =
(\rho \Gamma_d (d-1)!)^{-1/d} \log(\rho \Gamma_d L^d)$ exists, that governs
the behavior of average shortest-path distances as a function of Euclidean
separation.  This characteristic length diverges when $L \to \infty$, for any
$\rho>0$.  The typical separation $s=\rho^{-1/d}$ between shortcut
ends~\cite{newman2}, which is size-independent, is \emph{not relevant} for
$\ell(r)$. The validity of (\ref{eq:ellfinal}) has been verified numerically
in one dimension recently~\cite{corrlen}.

It is interesting to notice that the rescaled shortest-path distance $\tell=
(\rho \Gamma_d)^{1/d} \ell$ is a simple function of the rescaled Euclidean
distance $\terr= (\rho \Gamma_d)^{1/d} r$.
\begin{equation}
  \tell(\terr) = \left\{
    \begin{array}{lll}
      \terr &\hbox{for}  &\terr < 1  \\ 
      1     &\hbox{for}  &\terr \geq 1 
    \end{array} \right.
\label{eq:tell}
\end{equation}

Using (\ref{eq:ellfinal}) we can now calculate $\avell(\rho,L)$, the (global)
average shortest-path length~\cite{watts,bart,newman}, when $\tilL^d>>1$. One
has
\begin{eqnarray}
  \avell(\rho,L) &=& \frac{d}{L^{d}} \int_0^{L} \ell(r) r^{d-1} dr
  \nonumber \\ 
  &=& \frac{d}{L^{d}} \int_0^{r_c} r^d dr 
  +   \frac{r_c d}{L^{d}} \int_{r_c}^{L}
  r^{d-1} dr  \label{eq:first} \\
  &=& r_c \left[ 1 - \frac{1}{d+1} \left( \frac{r_c}{L} \right)^d \right]
\end{eqnarray}
So that the ``order parameter'' $\Orderpar=\avell/L$~\cite{firstorder,corrlen}
reads
\begin{equation}
\Orderpar = z ( 1 - \frac{z^d}{d+1}) 
\label{eq:universal}
\end{equation}
where $z=r_c/L$. 

When $\rho \to 0$ faster than $L^{1/d}$ (so that $\tilL^d <<1$), formula
(\ref{eq:first}) holds with $r_c \to L$, and thus $\Orderpar \to d/(d+1)$ as
expected.  On the other hand if $\rho > 0$ one has that $r_c << L$ when $L \to
\infty$, and thus $\Orderpar \to 0$ in this limit.  Therefore $\Orderpar$
undergoes a discontinuity at $\rho=0$ in the $L \to \infty$
limit~\cite{firstorder,corrlen}.

Notice that $r_c/L = \log(\rho \Gamma_d L^d)/L( \rho \Gamma_d (d-1)!)^{1/d}
\sim \log(L/s)/(L/s)$, , where $s \sim \rho^{-1/d}$~\cite{newman2} is the mean
separation between shortcut-ends. Thus $\Orderpar$ can be written as a
function of $L/s$ only. Therefore if we measure $\Orderpar$ on systems with
several values of $L$ and $\rho$ and plot the data versus $L/s$, we would find
that they \emph{collapse}~\cite{bart,barrat,newman,newman2}. Because of this
behavior some authors~\cite{newman,newman2} have suggested that the transition
at $\rho=0$ is a \emph{critical point} with a size-independent characteristic
length $\xi \sim s \sim \rho^{-1/d}$. Our results here and in previous
work~\cite{firstorder,corrlen} suggest that this is not the case.  According
to our calculation, the only characteristic length in regard to shortest-paths
is $r_c$, and it diverges with system size $L$.

We have thus shown that, on a finite system with $L^d p>>1$, two widely
separated timescales for spreading can be identified. The first one $t^* = (2p
\Gamma_d (d-1)!)^{-1/d}$ determines the crossover from normal (i.e.
proportional to $t^d$) to exponential spreading. A much larger timescale
$t_{sat}$ given by (\ref{eq:tsat}) determines the saturation of the spreading
process. This second timescale coincides with the lengthscale $r_c$ at which
the behavior of shortest path lengths $\ell(r)$ saturates, as given by Eq.
(\ref{eq:ellfinal}).

It is clear from our calculation that $r_c$ diverges with $L$ because the
locations of the secondary spheres are uncorrelated with the location of the
primary infection.  In other words, because on a system of size $L$, the
typical separation between both ends of a shortcut scales as $L$.  A different
situation would certainly arise if shortcuts had a length-dependent
distribution. For example one can connect each site $i$, with probability $p$,
to a single other site $j$, chosen with probability $r_{ij}^{-\alpha}$, where
$\alpha$ is a free parameter. For $\alpha \to 0$, this model is the same as
discussed here, while for $\alpha$ large one would only have short-range
connections and thus there would be no short-distance regime, even for $p=1$.
We are presently studying this general model~\cite{modified}.

\acknowledgments I acknowledge useful discussions with M.~Argollo de Menezes.
This work is supported by FAPERJ.


\begin{thebibliography}{99}
  
\bibitem{watts} D.~J.~Watts and S.~H.~Strogatz, ``Collective dynamics of
  small-world networks'', Nature {\bf 393}, 440 (1998).
  
\bibitem{herzel} H.~Herzel, ``How to quantify ``small world'' networks?'',
  Fractals {\bf 6}, 4 (1998).
  
\bibitem{pandit} S.~A.~Pandit and R.~E.~Amritkar, ``Characterization and
  control of small-world networks'', {\it chao-dyn/9901017}.
  
\bibitem{bart} M.Barth\'el\'emy and L.A.N.Amaral, ``Small-world networks:
  evidence for a crossover picture'', {\it cond-mat/9903108}, Phys. Rev. Lett.
  to appear.
  
\bibitem{barrat} A.~Barrat, ``Comment on `Small-world networks: evidence for a
  crossover picture'~ '', {\it cond-mat/9903323}.
  
\bibitem{monasson} R.~Monasson, ``Diffusion, localization and dispersion
  relations on small-world lattices'', {\it cond-mat/9903347}.
  
\bibitem{newman} M.~E.~J.~Newman and D.~J.~Watts, ``Renormalization group
  analysis of the small-world network model'', {\it cond-mat/9903357}.

\bibitem{weigt} A.~Barrat and M.~Weigt, ``On the properties of small-world
  network models'', {\it cond-mat/9903411}.
  
\bibitem{firstorder} M.~Argollo de Menezes, C.~Moukarzel and T.~J.~P.~Penna,
  ``First-order transition in small-world networks'', {\it cond-mat/9903426}. 

\bibitem{kasturi} R.~Kasturirangan, ``Multiple scales in small-world
  networks'', {\it cond-mat/9904055}.

\bibitem{newman2} M.~E.~J.~Newman and D.~J.~Watts, ``Scaling and percolation
  in the small-world network model'', {\it cond-mat/9904419}.
  
\bibitem{kulkarni} R.~V.~Kulkarni, E.~Almaas and D.~Stroud, ``Evolutionary
  dynamics in the Bak-Sneppen model on small-world networks'', {\it
    cond-mat/9905066}.
 
\bibitem{corrlen} C.~Moukarzel and M.~Argollo de Menezes, ``Infinite
  characteristic length in small-world systems'', {\it cond-mat/9905131}.
  
\bibitem{Stauffer} \textit{Introduction to Percolation Theory}, D.~Stauffer
  and A.~Aharony, 2nd ed. (Taylor \& Francis 1994).

\bibitem{Fractals} \textit{Fractals and Disordered Systems}, edited by A.~Bunde
  and S.~Havlin, 2nd ed. (Springer, Berlin, 1996).
    
\bibitem{modified} M.~Argollo de Menezes, C.~Moukarzel and T.~J.~P.~Penna, in
  preparation.

\end{thebibliography}
\end{document}